\documentclass[preprintnumbers,showkeys,superscriptaddress,showpacs,
nofootinbib,byrevtex,fleqn]{revtex4}
\usepackage{amsmath,amsfonts,amssymb,amscd,amsxtra,amsthm}
\usepackage{graphicx}
\usepackage{bm}
\usepackage{epstopdf}
\usepackage{multirow}
\begin{document}
\preprint{KIAS-P12005}
\preprint{INHA-NTG-07/2011}
\title{Generalized form factors and spin structures of the kaon}  
\author{Seung-il Nam}
\email[E-mail: ]{sinam@kias.re.kr}
\affiliation{School of Physics, Korea Institute for Advanced Study,
  Seoul 130-722, Korea} 
\affiliation{Research Institute for Basic Sciences, Korea Aerospace
University, Goyang  412-791, Korea}  
\author{Hyun-Chul Kim}
\email[E-mail: ]{hchkim@inha.ac.kr}
\affiliation{Department of Physics, Inha University, Incheon 402-751,
  Korea}   
\affiliation{School of Physics, Korea Institute for Advanced Study,
  Seoul 130-722, Korea}
\date{December, 2011}
\begin{abstract}
We investigate the spin structure of the kaon, based on the nonlocal
chiral quark model from the instanton vacuum. We first revisit the
electromagnetic form factors of the pion and kaon, improving the
results for the kaon. We evaluate the generalized tensor form factors
of the kaon in order to determine the probability density of
transversely polarized quarks inside the kaon. We consider the effects
of flavor SU(3) symmetry breaking, so that the  probability density of the
up and strange quarks are examined in detail. It is found that the
strange quark behaves differently inside the kaon in comparison with
the up quark.   
\end{abstract} 
\maketitle
\textbf{1.} Recently, the QCDSF/UKQCD Collaborations reported the
first lattice results for the pion
transversity~\cite{Brommel:2007xd}. They evaluated the probability
density of the polarized quarks inside the pion and found that their
distribution in the impact-parameter space is strongly distorted when
the quarks are transversely polarized. These lattice results have
triggered various theoretical 
works~\cite{Frederico:2009fk,Gamberg:2009uk,Broniowski:2010nt,Nam:2010pt}.    
Broniowski et al.~\cite{Broniowski:2010nt} have investigated the tensor
form factors of the pion within the local and nonlocal
chiral quark models~\cite{Broniowski:2010nt} and have employed a
larger value of the pion mass, i.e. $m_\pi = 600$ MeV in such a way
that the results can be compared with the lattice data directly. They
also considered the case of the chiral limit. In
Ref.~\cite{Nam:2010pt}, the tensor form factors of the pion were
calculated and the probability density of the polarized quarks inside
the pion was derived by combining the tensor form factors with the
electromagnetic (EM) ones, based on the nonlocal chiral quark model
(N$\chi$QM) from the instanton vacuum. The results were in good
agreement with the lattice data~\cite{Brommel:2007xd}. 

It is also of great interest to study the spin structure of the kaon, 
since it sheds light on the effects of flavor SU(3) symmetry breaking on 
behavior of the strange quark. Thus, in the present Letter, we aim at   
investigating the generalized tensor form factors of the kaon and
their implications for its spin structure within the framework of the
N$\chi$QM~\cite{Diakonov:1985eg,Musakhanov:1998wp}. In
Ref.~\cite{Musakhanov:1998wp}, Musakhanov extended the work of
Ref.~\cite{Diakonov:1985eg}, considering the finite current-quark 
mass. Since we need to take into account the explicit breaking of
flavor SU(3) symmetry breaking, we will use the extended N$\chi$QM
from the instanton
vacuum~\cite{Musakhanov:1998wp,Musakhanov:2001pc,Musakhanov:2002vu}.       
The model provides a suitable framework to study properties of the
kaon, since the instanton vacuum explains the spontaneous chiral
symmetry breaking (S$\chi$SB) naturally via quark zero modes and 
the explicit breaking of flavor SU(3) symmetry can be treated
consistently. Moreover, an important merit of this approach lies in
the fact that there are only two parameters, i.e. the average
(anti)instanton size $\bar{\rho}\approx1/3$ fm and average 
inter-instanton distance $\bar{R}\approx1$ fm. In particular, the
average size of instantons, of which the inverse is approximately
equal to $\bar{\rho}^{-1}\approx600$ MeV, provides the natural scale of the 
model. Note that the values of the $\bar{\rho}$ and $\bar{R}$ were  
estimated many years ago phenomenologically in
Ref.~\cite{Shuryak:1981ff} as well as theoretically in 
Refs.~\cite{Diakonov:1983hh,Diakonov:2002fq,Schafer:1996wv}. The
present framework has been already used to describe successfully
semileptonic decays of the kaon~\cite{Nam:2007fx} and qualitatively
its EM form factor~\cite{Nam:2007gf}.

In order to evaluate the probability density of the polarized quarks
inside the kaon, it is essential to know quantitatively the
EM form factor of the kaon. Thus, in the present work, we
will take as a free parameter the constituent-quark mass
at the zero virtuality of the quark rather than strictly following the
previous model derived from the instanton vacuum.  
We will show that while properties of the pion are very stable, those
of the kaon are much improved, compared to the results of the previous 
work~\cite{Nam:2007gf}. We will also calculate the tensor form factors
of the kaon with the two different values of the constituent-quark
mass. Then, the probability density of the polarized quarks 
inside the kaon will be derived and discussed, based on the results of
the EM and tensor form factors of the kaon. We will also study the
behavior of the up and strange quarks inside the kaon, so that we
understand the effects of flavor SU(3) symmetry breaking on the
distribution of the strange quarks inside the kaon.
\vspace{0.5cm}

\textbf{2.} The generalized tensor form factors $B_{ni}(Q^2)$ of the
pseudoscalar meson are defined as the matrix elements of the following
tensor operator 
\begin{equation}
\label{eq:BG}
\langle\phi(p_f)|\mathcal{O}^{\mu\nu\mu_1\cdots\mu_{n-1}}_T|\phi(p_i)\rangle
=\mathcal{AS}\left[\frac{(p^{\mu}q^{\nu}-q^{\mu}p^{\nu})}{m_{\phi}}
\sum^{n-1}_{i=\mathrm{even}}q^{\mu_1}\cdots q^{\mu_i}
p^{\mu_{i+1}}\cdots p^{\mu_{n-1}}B_{ni}(Q^2)\right],
\end{equation}
where $p_{i}$ and $p_{f}$ stand for the initial and final on-shell
momenta of  the pseudoscalar meson $\phi$, respectively. We assign the 
mass of the meson as $m_\phi$, and also introduce respectively the
average momentum and the momentum transfer $p=(p_{f}+p_{i})/2$ and
$q=p_{f}-p_{i}$. For definiteness, we will only take into account the  
positively charged pion $(\pi^+)$ and kaon $(K^+)$ for the meson in
this work. Moreover, we set the pion and kaon masses as $m_{\pi}=140$
MeV and $m_K=495$ MeV as numerical input. The tensor operator can
be expressed as 
\begin{equation}
\label{eq:OP}
\mathcal{O}^{\mu\nu\mu_1\cdots\mu_{n-1}}_T=
\mathcal{AS}
\left[\bar{\psi}_f\sigma^{\mu\nu}(i\tensor{D}^{\mu_1})
\cdots(i\tensor{D}^{\mu_{n-1}})\psi_f \right].
\end{equation}
The operations $\mathcal{A}$ and $\mathcal{S}$ denote the
anti-symmetrization in $(\mu,\nu)$ and symmetrization in
$(\nu,\cdots,\mu_{n-1})$ with the trace terms subtracted in all the
indices. The $\psi_f$ stands for a quark field with flavor $f$.

Taking into account Eqs.~(\ref{eq:BG}) and (\ref{eq:OP}), we can
define respectively the first and second generalized tensor form
factors $B^{\phi,f}_{10}$ and $B^{\phi,f}_{20}$ in the momentum space
as the matrix element of the tensor current, using the
auxiliary-vector method as in Ref.~\cite{Diehl:2010ru}:   
\begin{eqnarray}
\label{eq:TENSOR}
&&\langle \phi(p_{f})|\bar{\psi}_f(0)\sigma_{ab} \psi_f(0)
|\phi(p_{i})\rangle =
\left[(p_i\cdot a)(p_f\cdot b)-(p_i\cdot b)(p_f\cdot a)\right]
\frac{B^{\phi,f}_{10}(Q^{2})}{m_\phi},
\cr
&&\langle \phi(p_{f})|\bar{\psi}_f(0)
\sigma_{ab}(i\tensor{D}\cdot a) \psi_f(0)
|\phi(p_{i})\rangle =
\left\{(p\cdot a)[(p_i\cdot a)(p_f\cdot b)-(p_i\cdot b)(p_f\cdot a)]
\right\} \frac{B^{\phi,f}_{20}(Q^{2})}{m_\phi},
\end{eqnarray}
where the vectors satisfy the following conditions $a^2=a\cdot b=0$ 
and $b^2\ne0$, and we have used a shorthand notation 
$\sigma_{ab}\equiv\sigma_{\mu\nu}a^{\mu}b^{\nu}$. With the help of
this auxiliary-vector method, one can eliminate the trace-term 
subtractions. We also use the hermitized covariant derivative
$i\tensor{D}_\mu\equiv (i\roarrow{D}_\mu-i\loarrow{D}_\mu)/2$, where
$D_\mu$ indicates the SU($N_c$) covariant derivative. Since we  are
interested in the spatial distribution of the transversely  
polarized quark inside the meson, we need to consider the Fourier
transform of the form factors~\cite{Miller:2007uy}:    
\begin{equation}
\label{eq:FT}
\mathcal{F}^{\phi,f}(b^{2}_{\perp})=\frac{1}{(2\pi)^2}\int d^{2} q_{\perp}
e^{-ib_{\perp}\cdot q_{\perp}}\mathcal{F}^{\phi,f}(q^{2}_{\perp})
=\frac{1}{2\pi}\int^\infty_0 Q\,dQ\,J_0(bQ)\,\mathcal{F}^{\phi,f}(Q^{2}),
\end{equation}
where $\mathcal{F}^{\phi,f}$ designates a generic flavor form factor 
representing, for instance, $A^{\phi,f}_{n0}$ or $B^{\phi,f}_{n0}$. The
$b_{\perp}$ denotes the impact parameter that measures the distance
from the center of momentum of the meson to the quark in the
transversed plane to its motion. Here, we use $|q_\perp|\equiv Q$ and
$|b_\perp|\equiv b$. The $J_0$ stands for the Bessel function of order  
zero. Similarly, the Fourier transform of the derivative of the
generalized form factors with respect to $b^{2}_{\perp}$ can be
evaluated as 
\begin{equation}
\label{eq:DFT}
\frac{\partial \mathcal{F}^{\phi,f}(b^{2}_{\perp})}{\partial b^{2}_{\perp}}
\equiv [\mathcal{F}^{\phi,f}(b^{2}_{\perp})]'
=-\frac{1}{4\pi b}
\int^{\infty}_{0} Q^2\,dQ\,J_1(bQ)\,\mathcal{F}^{\phi,f}(Q^{2}).
\end{equation}
The $J_1$ denotes the Bessel function of order one. 

The probability density of the transversely polarized quark with
flavor $f$ is defined in terms of the generalized vector and tensor
form factors~\cite{Brommel:2007xd}:    
\begin{equation}
\label{eq:DENSITY}
\rho^{\phi,f}_{n}(b_{\perp},s_{\perp})=
\frac{1}{2}\left[A^{\phi,f} _{n0}(b^{2}_{\perp})
-\frac{s^{i}_{\perp}\epsilon^{ij}b^{j}_{\perp}}{m_{\phi}}
\frac{\partial B^{\phi,f} _{n0}(b^{2}_{\perp})}{\partial b^{2}_{\perp}}\right],
\end{equation}
where the $s_\perp$ stands for the fixed transverse spin of the
quark. For simplicity, we choose the $z$ direction for the quark
longitudinal momentum. In the case of exact flavor SU(3) symmetry, the vector
form factor $A^{\phi,f} _{10}$ with flavor $f$ is just equal to the EM
form factor due to isospin symmetry~\cite{Brommel:2006ww}: $A^{\pi,u} 
_{10}(Q^2)=-A^{\pi,d} _{10}(Q^2)=F_{\pi}(Q^2)$. Similarly, we have the
following relation for the kaon: $A^{K,u} _{10}(Q^2)=-A^{K,s}
_{10}(Q^2)=F_{K}(Q^2)$ because of $V$-spin symmetry. However, these
simple relations are broken by explicit flavor SU(3) symmetry
breaking. Hence, it is necessary to compute separately the up-down and
strange form factors of the kaon. 

In the previous works~\cite{Nam:2007gf,Nam:2010pt}, the 
positively charged pion and kaon electromagnetic form factors were
already investigated. Thus, we will employ the same theoretical
framework to compute the vector form factors. Although the kaon vector
form factors are different from the EM form factor of the kaon as 
mentioned above, they can be easily evaluated within the same
framework. Therefore, we will only focus on how to evaluate 
the tensor form factors. For more details, one refers 
to Ref.~\cite{Nam:2007gf}. Considering all the ingredients discussed 
so far, one is finally led to the analytical definition of the
flavor probability density of the quark in the transverse
impact-parameter space as follows:    
\begin{equation}
\label{eq:DENSITY2}
\rho^{\phi,f}_n\left(b_{\perp},s_{x}=\pm 1\right)=
\frac{1}{2}\left[A^{\phi,f} _{n0}(b^{2})
\mp\frac{b\sin\theta_\perp}{m_{\phi}}[B^{\phi,f} _{n0}(b^{2})]'\right],
\end{equation}
where the spin of the quark inside the meson is quantized along the $x$
axis, $s_{\perp}=(\pm1,0)$.   
\vspace{0.5cm}

\textbf{3.} We now briefly explain the extended N$\chi$QM from the
instanton
vacuum~\cite{Musakhanov:1998wp,Musakhanov:2001pc,Musakhanov:2002vu}
and derive the generalized form factors of the pion and
kaon. Considering first the dilute instanton liquid, characterized by
two instanton parameters, i.e. the average (anti)instanton size
$\bar{\rho}\approx1/3$ fm and average inter-instanton distance
$\bar{R}\approx1$ fm with the small packing parameter
$\pi\bar{\rho}^4/\bar{R}^4\approx 0.1$, we are able to 
average the fermionic determinant over collective coordinates of
instantons with fermionic quasi-particles, i.e. the constituent quarks
introduced. The averaged determinant is reduced to the light-quark
partition function that can be given as a functional of the tensor
field in the present case.  Having bosonized and integrated it over
the quark fields, we obtain the following effective nonlocal
chiral action in the large $N_c$ limit in Euclidean space: 
\begin{equation}
\label{eq:ACTION1}
\mathcal{S}_{\mathrm{eff}}[m_f,\phi]=-\mathrm{Sp}
\ln\left[i\rlap{/}{\partial}+im_f+i\sqrt{M(\partial^2)}U^{\gamma_5}(\phi)
\sqrt{M(\partial^2)}+\sigma\cdot T \right],
\end{equation}
where $m_f$, $\phi$, and $\mathrm{Sp}$ indicate the current-quark mass,
the pseudo-Nambu-Goldstone (NG) boson field, and the functional
trace over all relevant spaces, respectively. Assuming isospin
symmetry and explicit flavor SU(3) symmetry breaking, we use the following
numerical values $m_{u}=m_d=5$ MeV and $m_s=180$ MeV. The
$M(i\partial)$ stands for the momentum-dependent effective quark mass,
generated from the fermionic zero modes of the
instantons~\cite{Diakonov:1985eg}. Its analytical form is in
general given by
\begin{equation}
\label{eq:MMMD}
M(\partial^2)=M_0F^2(t),\,\,\,\,
F(t)=2t\left[I_0(t)K_1(t)-I_1(t)K_0(t)-\frac{1}{t}I_1(t)K_1(t) \right].
\end{equation}
Here, $t=|\rlap{/}{\partial}|\bar{\rho}/2$, and $I_n$ and $K_n$
stand for the modified Bessel functions with the order $n$. In the
numerical calculations, instead of using Eq.~(\ref{eq:MMMD}), we will
make use of the following parametrization for numerical convenience: 
\begin{equation}
\label{eq:EFFM}
M(\partial^2)=M_{0}
\left(\frac{2}{2+\bar{\rho}^{2}\partial^{2}} \right)^{2},
\end{equation}
where $M_{0}$ indicates the constituent-quark mass at zero quark
virtuality and its value is determined by the self-consistent equation
of the instanton model in the chiral
limit~\cite{Musakhanov:1998wp,Musakhanov:2001pc,Musakhanov:2002vu}: 
\begin{equation}
\label{eq:SELF}
\frac{1}{\bar{R}^4}=4N_c\int\frac{d^4p}{(2\pi)^4}\frac{M^2(p)}{p^2+M^2(p)},
\end{equation}
resulting in $M_0\approx350$ MeV. As done in
Refs.~\cite{Musakhanov:1998wp,Musakhanov:2001pc,Nam:2006ng}, $M_0$ is 
modified due to the explicit flavor symmetry breaking in the following
way: 
\begin{equation}
\label{eq:POB}
M_0\to M_0\,f(m_f)=M_0\left[\sqrt{1+\frac{m^2_f}{d^2}}-\frac{m_f}{d}\right],
\end{equation}
where $d=198$ MeV . We will call
this model with the original set of parameters as {\it model I}. Note
that $M_0$ should be modified in such a way that the instanton-number
density is independent of the finite current-quark 
mass~\cite{Musakhanov:1998wp,Musakhanov:2001pc,Nam:2006ng}. On the
other hand, considering theoretical uncertainties in the instanton
framework for the flavor SU(3) sector, we still can choose the value
of $M_0$ to reproduce experimental data such as the pion and kaon
electric-charge radii, setting $f(m_f)=1$ in Eq.~(\ref{eq:POB}), from
a very phenomenological point of view, whereas the instanton
parameters remain unchanged. We call this phenomenological way as
{\it model II}. The pseudo-NG boson field is represented in a  
nonlinear form as~\cite{Diakonov:1995qy}:     
\begin{equation}
\label{eq:CHIRALFIELD}
U^{\gamma_5}(\phi)=
\exp\left[\frac{i\gamma_{5}(\lambda\cdot\phi)}{F_{\phi}}\right]
=1+\frac{i\gamma_{5}(\lambda\cdot\phi)}{F_{\phi}}
-\frac{(\lambda\cdot\phi)^{2}}{2F^{2}_{\phi}}+\cdots,
\end{equation}
where $\phi^{\alpha}$ is the flavor SU(3) multiplet defined as
\begin{equation}
\label{eq:PHI}
\lambda\cdot\phi=\left(
\begin{array}{ccc}
\frac{\pi^{0}}{\sqrt{2}}+\frac{\eta}{\sqrt{6}}&\pi^{+}&K^+\\
\pi^-&-\frac{\pi^{0}}{\sqrt{2}}+\frac{\eta}{\sqrt{6}}&K^0\\
K^-&\bar{K}^0&-\frac{2\eta}{\sqrt{6}}\\
\end{array}
 \right),
 \end{equation}
where the trace over the isospin space is defined by
$\mathrm{tr}[\lambda^{\alpha}\lambda^{\beta}]=2\delta^{\alpha\beta}$. The
$F_{\phi}$ denotes the weak-decay constant for the pseudo-NG bosons,
whose empirical values are $93.2$ MeV for the pion and $113$ 
MeV for the kaon for instance. The last term in Eq.~(\ref{eq:ACTION1})
denotes $\sigma\cdot T=\sigma_{\mu\nu} T_{\mu\nu}$, where
$\sigma_{\mu\nu}=i[\gamma_\mu,\,\gamma_\nu]/2$ and 
$T_{\mu\nu}$ designates the external tensor field. 

The three-point correlation function in Eq.~(\ref{eq:TENSOR}) can be
easily calculated by a functional differentiation with respect to the
pseudo-NG boson and external tensor fields. Having performed the
functional trace and that over color space, we can write the  
matrix elements for the $B^{\phi,f}_{10}(Q^2)$ and
$B^{\phi,q}_{20}(Q^2)$, corresponding to Eq.~(\ref{eq:TENSOR}), as
follows:  
\begin{eqnarray}
\label{eq:MAT2}
\langle \phi(p_{f})|q^{\dagger}(0)\sigma_{ab}q(0)
|\phi(p_{i})\rangle
&=& -\frac{2N_{c}}{F^{2}_{\phi}}
\int\frac{d^4k}{(2\pi)^4}
\mathrm{Tr}_{\gamma}
\left[
\frac{1}{i\rlap{\,/}{D}_{1}}\sqrt{M_{1}}\gamma_{5}\sqrt{M_{2}}
\frac{1}{i\rlap{\,/}{D}_{2}}\sqrt{M_{2}}\gamma_{5}\sqrt{M_{3}}
\frac{1}{i\rlap{\,/}{D}_{3}}\sigma_{ab}
\right],
\cr
\langle \phi(p_{f})|q^{\dagger}(0)\sigma_{ab}(i\tensor{D}\cdot a)q(0)
|\phi(p_{i})\rangle &=&-\frac{2N_{c}}{F^{2}_{\phi}}
\int\frac{d^4k}{(2\pi)^4}
\mathrm{Tr}_{\gamma}
\left[
\frac{1}{i\rlap{\,/}{D}_{1}}\sqrt{M_{1}}\gamma_{5}\sqrt{M_{2}}
\frac{1}{i\rlap{\,/}{D}_{2}}\sqrt{M_{2}}\gamma_{5}\sqrt{M_{3}}
\frac{1}{i\rlap{\,/}{D}_{3}}\sigma_{ab}\eta\right],
\end{eqnarray}
where $\eta\equiv (k+\frac{p_i}{2})\cdot a$. The relevant momenta are also defined as    
\begin{eqnarray}
\label{eq:MOM}
k_{1}&=&k-\frac{p_{i}}{2}-\frac{q}{2},
\,\,\,\,
k_{2}=k+\frac{p_{i}}{2}-\frac{q}{2},
\,\,\,\,
k_{3}=k+\frac{p_{i}}{2}+\frac{q}{2}.
\end{eqnarray}
Here, we have used the notation $M_{i}\equiv M(k^2_{i})$ for
$i=(1,2,3)$. The denominators become $\rlap{\,/}{D}_{i} =
\rlap{/}{k}_{i} + i\bar{M}_{i}$ in Eq.~(\ref{eq:MAT2}), where
$\bar{M}_{i}=M_{i}+m_{i}$. For $B^{K,u}_{n0}$,  we choose
$m_{1,3}=m_u$ and $m_2=m_s$, while we set $m_{1,3}=m_s$ and $m_2=m_u$
for $B^{K,s}_{n0}$. In order to evaluate the matrix element, we define
the initial and final pion momenta in the Breit (brick-wall) frame in
Euclidean space as done in Ref.~\cite{Nam:2007gf}. We also have chosen
the auxiliary vectors explicitly as $a=(0,1,0,i)$ and $b=(1,0,1,0)$,
which satisfy the conditions mentioned previously, and have defined
$\sigma_{ab}=\sigma_{\mu\nu}a^\mu b^\nu$. The momentum-dependent
effective quark mass $M_{a,b,c}$ can be also defined by using
Eqs.~(\ref{eq:EFFM}) and (\ref{eq:MOM}).  
\vspace{0.5cm}

\begin{figure}[t]
\begin{tabular}{cc}
\includegraphics[width=8.5cm]{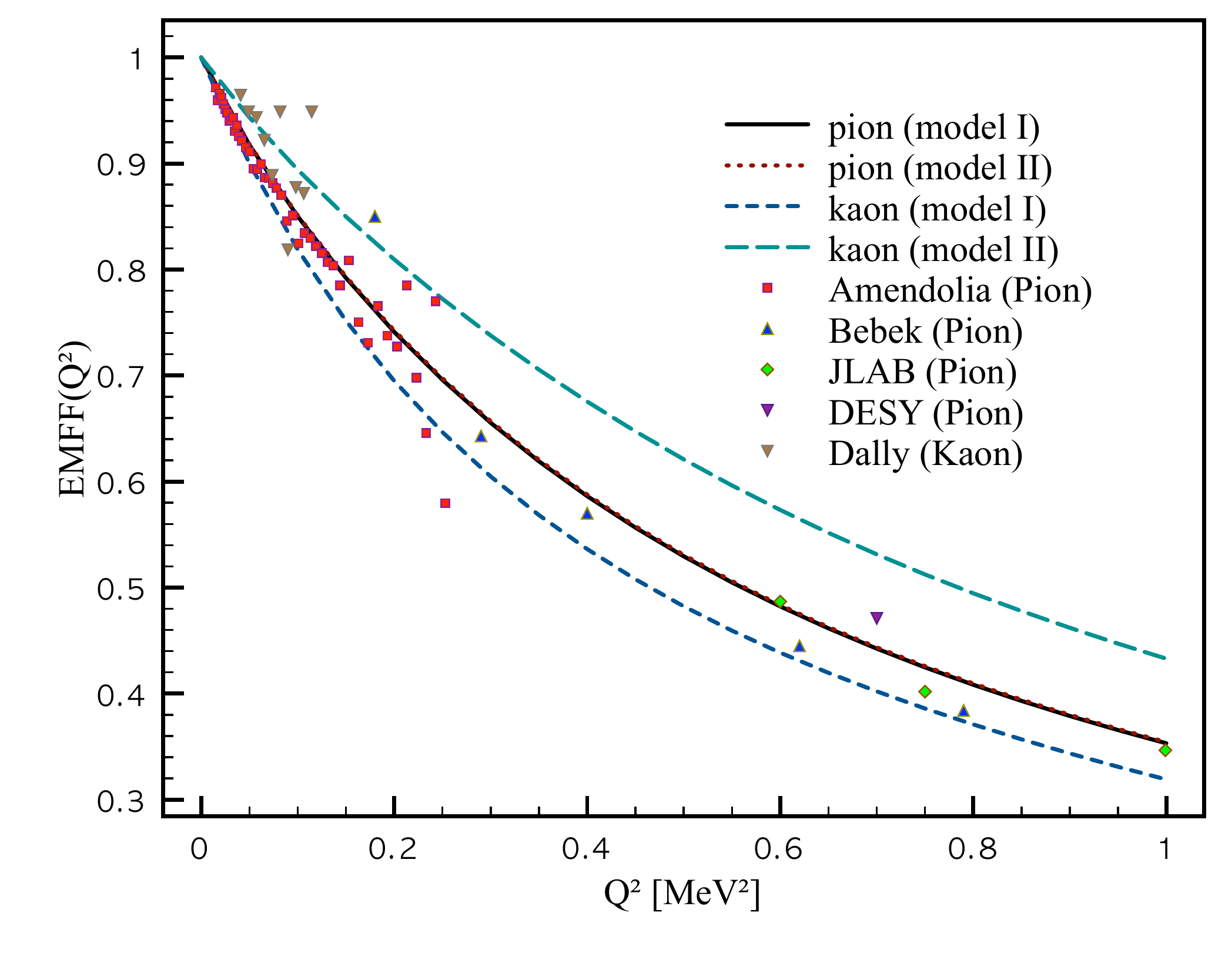}
\includegraphics[width=8.5cm]{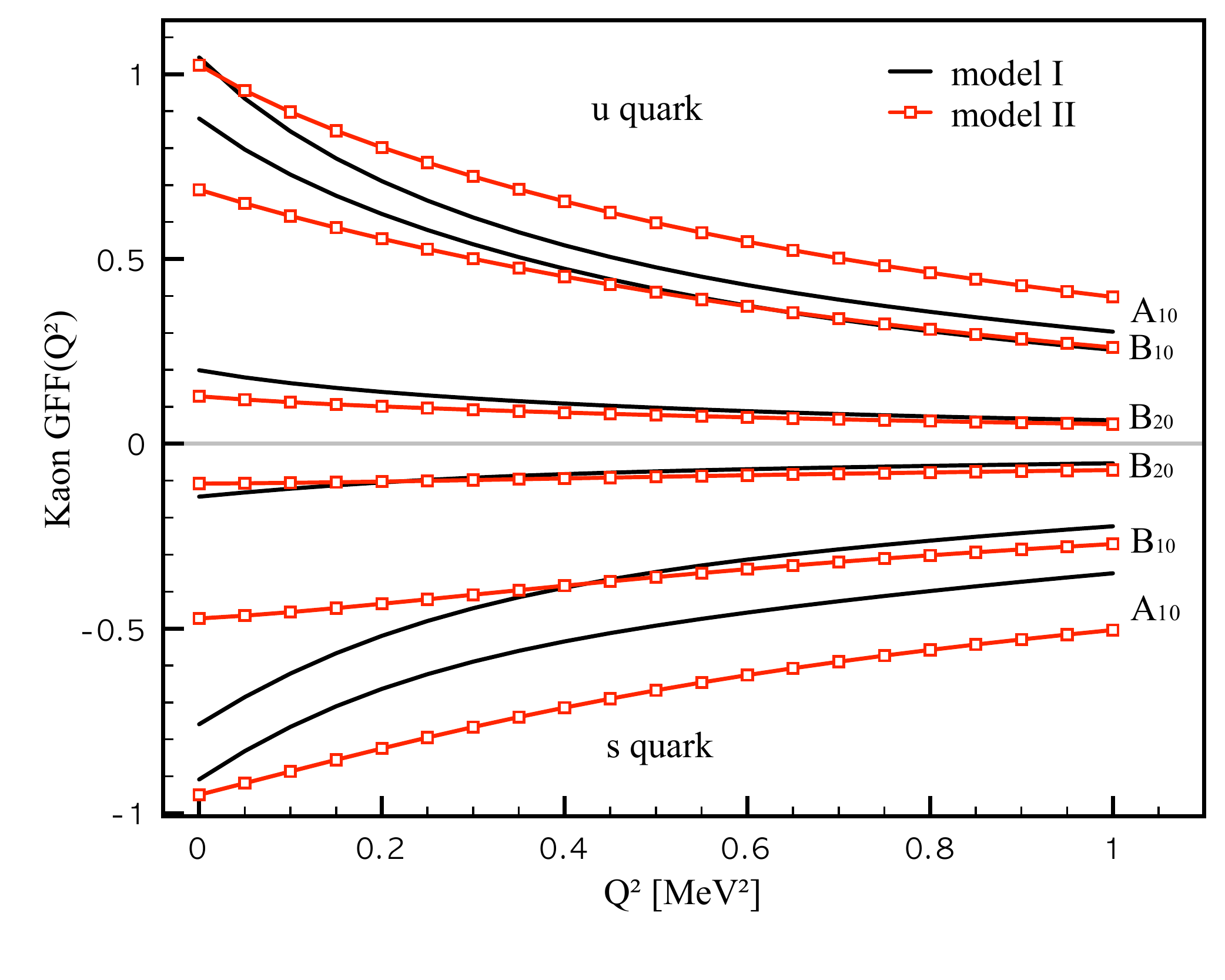}
\end{tabular}
\caption{(Color online) Electromagnetic form factors of the pion and
  kaon from model I and model II as functions of $Q^2$ in  
  the left panel. The experimental data are taken from
  Refs.~\cite{Amendolia:1986wj,Amendolia:1986ui,Bebek:1974iz,
Volmer:2000ek,Brauel:1979zk,Dally:1980dj}. Kaon
  generalized form factors, $A^{K,f}_{10}$, $B^{K,f}_{10}$, and 
  $B^{K,f}_{20}$, as functions of $Q^2$ for each flavor, based on 
  model I (solid) and model II (square) in the right panel. }        
\label{FIG1}
\end{figure}

\textbf{4.} We now present the numerical results and discuss
them. First, in order to see the reliability of the present framework,
we have computed the positive-charged pion and kaon 
EM form factors. Moreover, the vector form factors can be easily
derived from them. In the left panel of Figure~\ref{FIG1},
we draw the numerical results for the EM form factors of the pion and
kaon, using model I and model II, separately. Note that the results of 
model I are the same as those given in Ref.~\cite{Nam:2007gf} 
as they should be. As for the pion, the results from the two models
almost coincide with each other because of the small masses of the
light quarks in comparison to the renormalization point
$\mu\simeq600$ MeV, and turn out to be in good agreement with the
experimental data taken from 
Refs.~\cite{Amendolia:1986wj,Amendolia:1986ui,Bebek:1974iz,
Volmer:2000ek,Tadevosyan:2007yd, Brauel:1979zk}. This
renormalization-point value is proportional to the inverse of the
average (anti)instanton size,
i.e. $\mu\approx1/\bar{\rho}$~\cite{Shuryak:1981ff,Diakonov:1983hh,
Diakonov:2002fq,Schafer:1996wv}, indicating the scale of the
quark-(anti)instanton interaction strength. Though model I 
provides considerably good results, we made fine-tuning of the value
of $M_0$ to be $343$ MeV for model II to fit the electric-charge
radius of the pion $\langle
r^2\rangle^\mathrm{exp.}_\pi\approx(0.672\,\mathrm{fm})^2$. The pion 
decay constant is reproduced to be $F_\pi\approx94$ MeV for both
models, while its empirical value is $F_\pi=93.2$ MeV. All the
numerical results are listed in Table~\ref{TABLE0}.  

In contrast, the kaon EM form factors depend on which model we are
using. The results of model I are underestimated, being compared with
the experimental data~\cite{Dally:1980dj}: For example, the kaon
charge radius turns out to be $\langle
r^2\rangle^\mathrm{theo.}_K=(0.639\,\mathrm{fm})^2$, which is about 
$10\%$ larger than the experimental one $\langle
r^2\rangle^\mathrm{exp.}_K=(0.560\,\mathrm{fm})^2$. In addition, we 
have the smaller value of the kaon decay constant $F_K=100.17$ MeV in
comparison to the corresponding empirical value $F_K=113$ MeV. These
sizable deviations from the data have been 
already observed in our previous work~\cite{Nam:2007gf}, and can be
understood by the absence of the meson-loop corrections (MLC) which
is essential for the cases with the explicit flavor SU(3) symmetry
breaking~\cite{Nam:2008bq}. Hence, to remedy this problem for 
the kaon case, one may consider the MLC coming from the mesonic
fluctuations around the saddle point. However, since it is rather
involved, we take a more phenomenological stand on this problem. So, 
model II can be regarded as a phenomenological way in confronting with 
the experimental data. We now fit the value of $M_0$ to   
reproduce the experimental value for the kaon electric charge
radius. The fitted value $M_0=407$ MeV brings also about the kaon
decay constant $F_K=118.64$ MeV which is in good agreement with the
data. We note that the nonlocal contributions to the electromagnetic
form factors of the pion and kaon, which were already discussed in
Ref.~\cite{Nam:2007gf}, are about $(30-40)\,\%$. All the numerical
results can be found in Table~\ref{TABLE0}.      
\begin{table}[htb]
\begin{tabular}{c|c|c|cc|cc} \hline\hline
\,\,\,\,$\phi$\,\,\,\,& model 
&\hspace{0.5cm}$M_0$\hspace{0.5cm}
&\hspace{0.5cm}$F_\phi$\hspace{0.5cm}
&\hspace{0.5cm}$F^\mathrm{phen.}_\phi$\hspace{0.5cm}
&\hspace{0.5cm}$\langle r^2\rangle^\mathrm{theo.}_\phi$\hspace{0.5cm}
&\hspace{0.5cm}$\langle r^2\rangle^\mathrm{exp.}_\phi$\hspace{0.5cm}\\
\hline
\multirow{2}{*}{$\pi$}&I
&$350$ MeV
&$94.23$ MeV
&\multirow{2}{*}{$93.2$ MeV}
&$(0.673\,\mathrm{fm})^2$
&\multirow{2}{*}{$[(0.672\pm0.008)\,\mathrm{fm}]^2$}\\
&II
&$343$ MeV
&$94.43$ MeV
&
&$(0.672\,\mathrm{fm})^2$
&\\
\hline
\multirow{2}{*}{$K$}
&I
&$350$ MeV
&$100.17$ MeV
&\multirow{2}{*}{$113$ MeV}
&$(0.639\,\mathrm{fm})^2$
&\multirow{2}{*}{$[(0.560\pm0.031)\,\mathrm{fm}]^2$}\\
&II
&$407$ MeV
&$118.64$ MeV
&
&$(0.560\,\mathrm{fm})^2$
&\\
\hline
\hline
\end{tabular}
\caption{The results of the decay constants and charge radii of the
  pion and kaon from model I and model II, respectively. We have used 
  $m_{u,d}=5$ MeV, $m_s=180$ MeV, 
  $1/\bar{R}^4=(200\,\mathrm{MeV})^4$, and $1/\bar{\rho}=600$ MeV.} 
\label{TABLE0}
\end{table}

Once all the relevant parameters are fixed as discussed above
(see Table~\ref{TABLE0}), we can proceed to compute the generalized
form factors of the kaon within the two models. As for the pion form
factors, one can refer to our previous 
work~\cite{Nam:2010pt}. In the left panel of Figure~\ref{FIG1}, we
depict the results from model I in solid (pion) and shot-dashed (kaon)
curves, where as those from model II are drawn in dotted (pion) and
long-dashed (kaon) curves. Being different from the pion case, 
it is necessary to consider the up and strange quarks separately in
Eq.~(\ref{eq:OP}). However, the up and strange vector form factors of
the kaon must satisfy the charge conservation at $Q^2=0$ as follows: 
\begin{equation}
\label{eq:}
\frac{2}{3}A^{K,u}_{10}(0)-\frac{1}{3}A^{K,s}_{10}(0)=1.
\end{equation}
In the right panel of Figure~\ref{FIG1}, we show the results of the up
and strange generalized vector and tensor form factors. The results
from model I are drawn in solid curves, whereas those from model II
are distinguished by putting squares. The six curves in the upper part 
correspond to the up form factors and those in the lower part
illustrate the strange ones. One can observe from the numerical
results that the difference between the results from the two different
models is noticeable in general apart from the $B_{20}$. 
Before we derive the probability densities of the polarized quarks, it
is very convenient to parameterize the form
factors~\cite{Brommel:2007xd}:  
\begin{equation}
\label{eq:PARAFF}
A^{\phi,f}_{10}(Q^2)\to
\frac{A^{\phi,f}_{10}(0)}{1+Q^2/M^2_{A^{\phi,f}_{10}}},\,\,\,\,
B^{\phi,f}_{n0}(Q^2)\to
\frac{B^{\phi,f}_{n0}(0)}{[1+Q^2/(p_nM^2_{B^{\phi,f}_{n0}}) ]^{p_n}},
\end{equation}
where the $M_{A^{K,f}_{n0}}$ and $M_{B^{K,f}_{n0}}$ denote the pole
masses corresponding to the flavor form factors. Note that we employ a 
simple monopole and $p$-pole type parametrizations for the vector and
tensor form factors, respectively. Considering the condition $p>1.5$
for the regular behavior of the probability density at
$b_{\perp}\to0$~\cite{Diehl:2005jf} and following 
Ref.~\cite{Brommel:2007xd}, we take $p_{1}=p_{2}=1.6$ as a
trial. These parametrized form factors are also very useful in
analyzing the lattice simulation~\cite{Brommel:2007xd}. Using the
numerical results for the generalized form factors drawn in the left
panel of Figure~\ref{FIG1} and Eq.~(\ref{eq:PARAFF}), we extract the
numerical values for the parametrized form factors as
\begin{eqnarray}
(M_{A^{K,u}_{10}},M_{B^{K,u}_{10}},M_{B^{K,u}_{20}})&=&
(0.647,\,0.726,\,0.748)\,\mbox{GeV},\cr
 (M_{A^{K,s}_{10}},M_{B^{K,s}_{10}},M_{B^{K,s}_{20}}) &=&
 (0.772,\,0.709,\,0.806)\, \mbox{GeV}   
\end{eqnarray}
for model I and 
\begin{eqnarray}
(M_{A^{K,u}_{10}},M_{B^{K,u}_{10}},M_{B^{K,u}_{20}}) &=&
(0.894,\,0.898,\,0.918) \,\mbox{GeV},\cr 
(M_{A^{K,s}_{10}},M_{B^{K,s}_{10}},M_{B^{K,s}_{20}}) &=&
(1.081,\,1.272,\,1.520) \,\mbox{GeV}  
\end{eqnarray}
for model II. The results of the pole masses from model II turn out to
be in general larger than those from model I, which indicates that the
the form factors from model II decrease less slowly than those from
model I, as shown in Figure~\ref{FIG1}. The pole masses of the pion
tensor form factors ~\cite{Nam:2010pt} $M_{B^{\pi,u}_{10}}=0.761$ GeV
and $M_{B^{\pi,u}_{20}}=0.864$ GeV~\footnote{We note that there was an
  error in Ref.~\cite{Nam:2010pt} related to the tadpole
  diagram. The contribution from this diagram is essentially zero
  for the tensor form factors due to its antisymmetric
  nature. However, correcting the error brings about 
  negligible changes in the numerical results.} can be compared with
those from the lattice simulation, $(0.756\pm0.095)$ GeV and
$(1.130\pm0.265)$ GeV, respectively~\cite{Brommel:2006ww}. Note that
these lattice data are extrapolated values to $m_\pi=140$ MeV from the
higher pion mass $m_\pi\approx600$ MeV. In
Ref.~\cite{Dorokhov:2011ew}, the Holdom-Terning-Verbeek (HTV)
nonlocal-interaction model was employed to compute the tensor form
factors for the pion. Their results are qualitatively compatible with
ours given in Ref.~\cite{Nam:2010pt}. All the numerical results are
summarized in Table~\ref{TABLE1} in addition to the values of the  
generalized form factors at $Q^2=0$.   
\begin{table}[hb]
\begin{tabular}{c|c|cc|cc|cc|c} \hline\hline
Model&Quark&\,\,\,\,$A^{K,q}_{10}(0)$\,\,\,\,
&$M_{A^{K,q}_{10}}$
&\,\,\,\,$B^{K,q}_{10}(0)$\,\,\,\,
&$M_{B^{K,q}_{10}}$
&\,\,\,\,$B^{K,q}_{20}(0)$\,\,\,\,
&$M_{B^{K,q}_{20}}$&\,\,\,\,$\langle b^{K,q}_y\rangle$\,\,\,\,\\
\hline
\multirow{2}{*}{I}
&$q=u$
&$1.045$&$0.647$ GeV
&$0.880$&$0.726$ GeV
&$0.199$&$0.748$ GeV
&$0.168$ fm\\
&$q=s$
&$-0.909$&$0.772$ GeV
&$-0.760$&$0.709$ GeV
&$-0.143$&$0.806$ GeV
&$0.166$ fm\\
\hline
\multirow{2}{*}{II}&$q=u$
&$1.025$&$0.894$ GeV
&$0.687$&$0.898$ GeV
&$0.128$&$0.918$ GeV
&$0.133$ fm\\
&$q=s$
&$-0.950$&$1.081$ GeV
&$-0.473$&$1.272$ GeV
&$-0.108$&$1.520$ GeV
&$0.100$ fm\\
\hline
\hline
\end{tabular}
\caption{Results for the parametrized form factors in
  Eq.~(\ref{eq:PARAFF}) and $\langle b^{K,q}_y\rangle$ for model I
  and model II.} 
\label{TABLE1}
\end{table}

\begin{figure}[ht]
\begin{tabular}{cc}
\includegraphics[width=9cm]{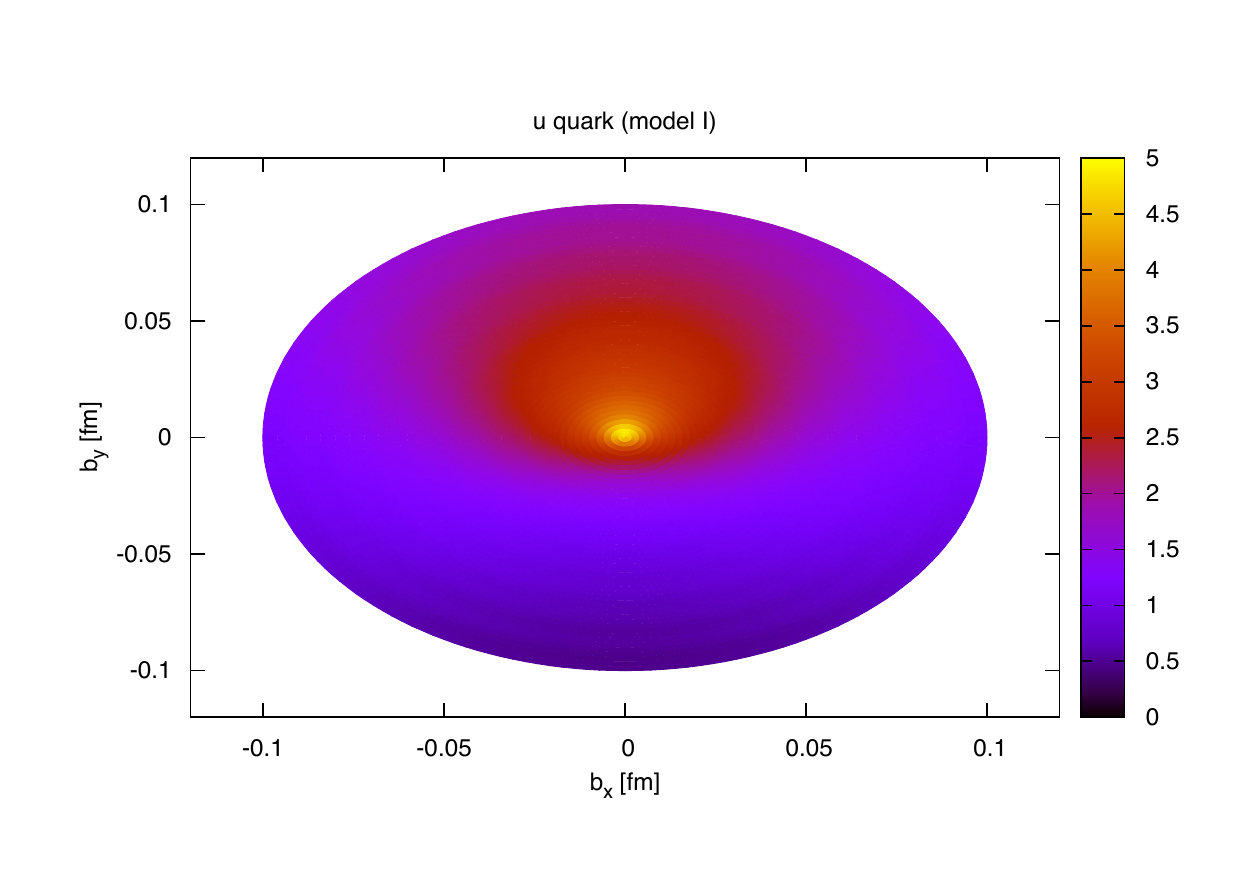}
\includegraphics[width=9cm]{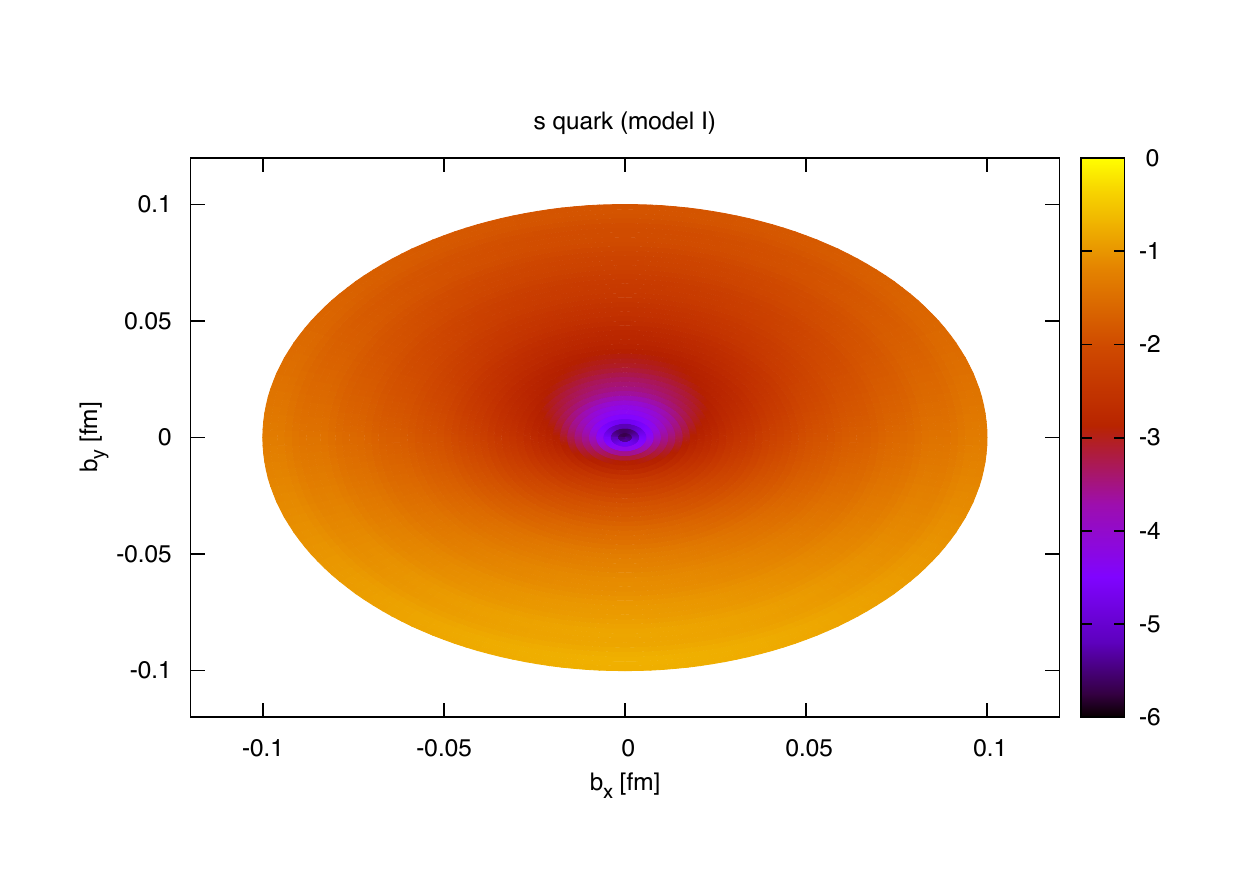}
\end{tabular}
\begin{tabular}{cc}
\includegraphics[width=9cm]{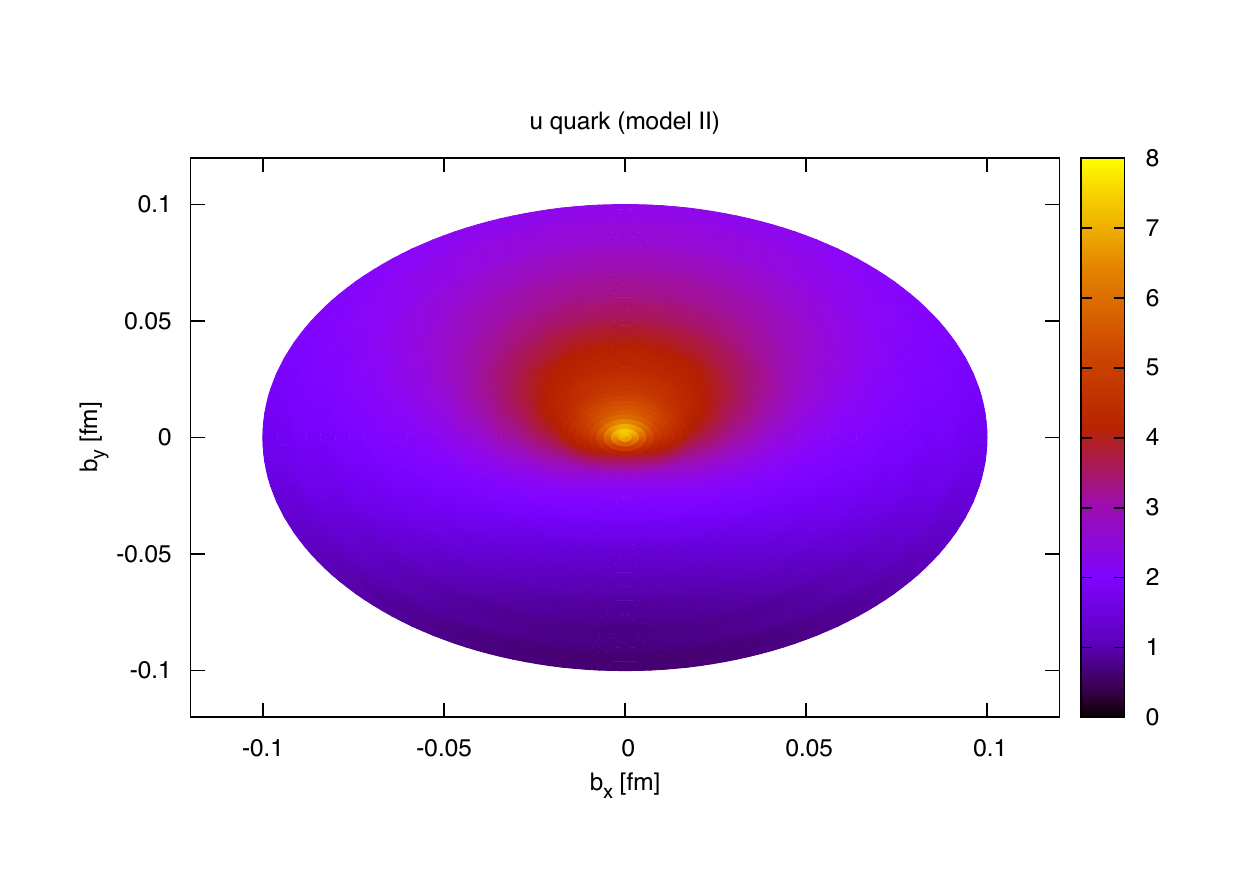}
\includegraphics[width=9cm]{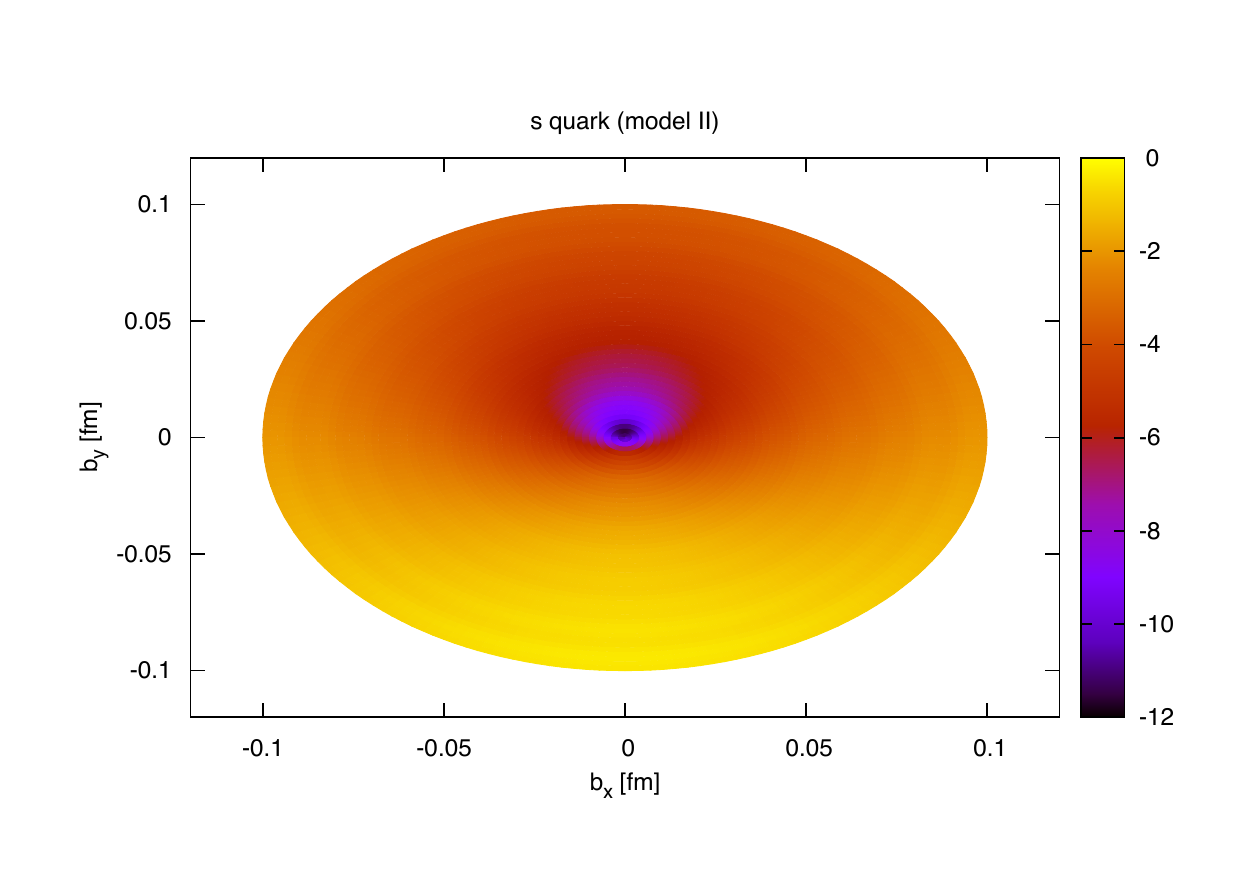}
\end{tabular}
\caption{(Color online) Polarized quark-spin density $\rho^{K,f}_1$
  for the up (left   column) and strange (right column) quarks as
  functions of the impact parameters $b_x$ and $b_y$ for model I
  (upper panels) and II (lower panels).}        
\label{FIG2}
\end{figure}

We are now in a position to consider the quark-spin probability
density $\rho^{\phi,q}_n$, defined in Eq.~(\ref{eq:DENSITY}), using
our numerical results for the generalized form factors. For
definiteness, we choose $s_\perp=+1$ explicitly in
Eq.~(\ref{eq:DENSITY2}) and take the absolute values 
for the densities. After performing the Fourier transform of the form
factors, we show the numerical results as functions of the
two-dimensional impact-space parameters, i.e. $b_x$ and $b_y$ as shown
in Figure~\ref{FIG2}. As already shown in Ref.~\cite{Nam:2010pt} and
understood from Eq.~(\ref{eq:DENSITY2}), the unpolarized-quark
probability density $(s_\perp=0)$ must be symmetric under the rotation
with respect to the $z$ axis, being perpendicular to the $b_x$-$b_y$
plane. Hence, we do not see any interesting structures from them. On
the contrary, when the quark inside the meson is polarized
($s_\perp=+1$), there appears tilted structures signaling the spin
structure inside the meson. 
In Figure~\ref{FIG2}, we show the numerical results
for the polarized densities from model I (upper-two panels) and from
model II (lower-two panels). In the left (right) column, we depict
them for the up (strange) quarks. One can clearly observe the
distortions of the surfaces due to the polarization. 
\begin{figure}[ht]
\includegraphics[width=9cm]{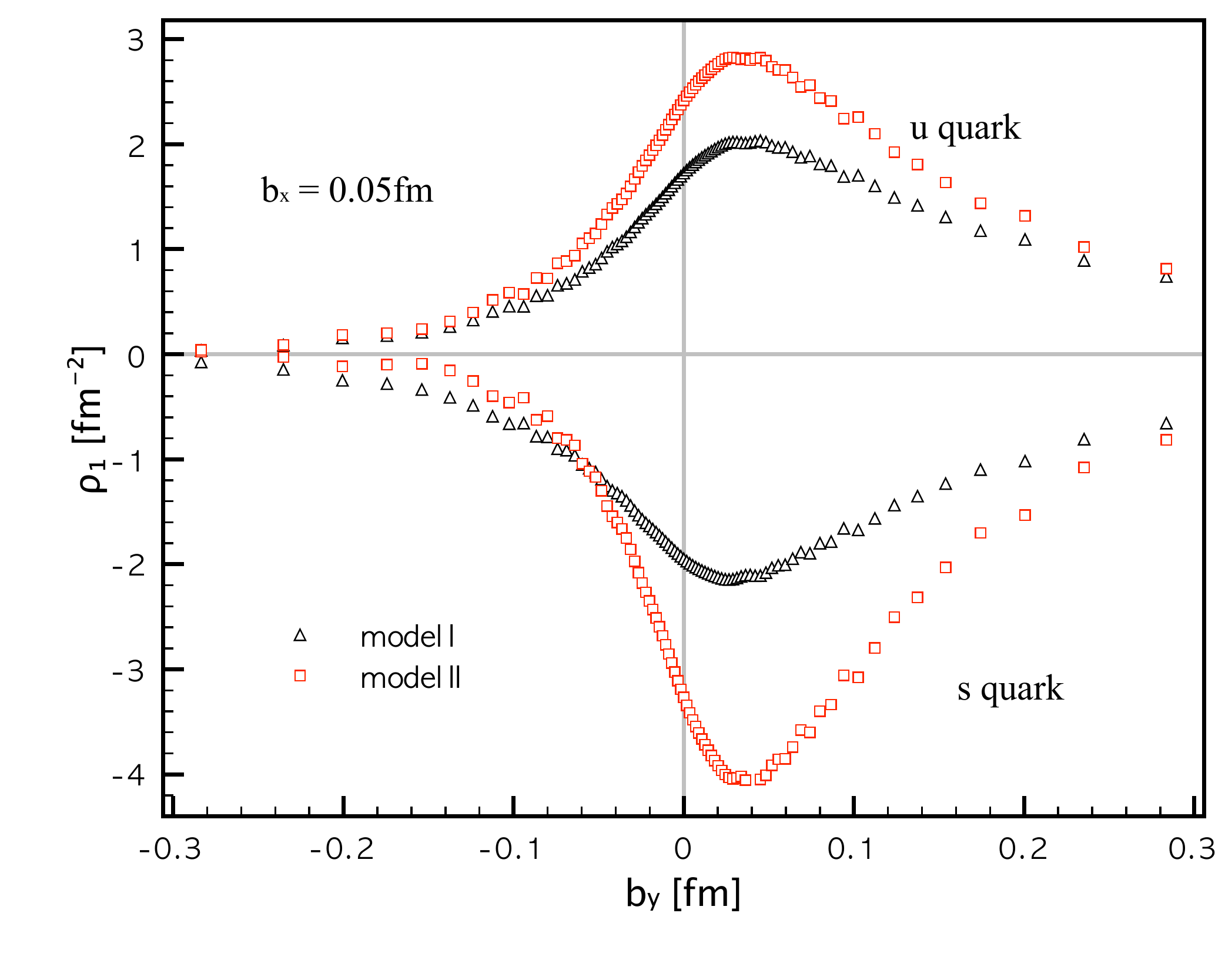}
\caption{(Color online) Profiles of the probability densities at
  $b_x=0.05$ fm for the $u$-quark (upper-two curves) and $s$-quark
  (lower-two curves)  for the model I (solid) and II (square).}        
\label{FIG3}
\end{figure}
For instance, in 
Figure~\ref{FIG3}, we draw the profiles of probability densities of the
polarized up and strange quarks as functions of $b_y$ at $b_x=0.05$
fm, separately for each model. While the peak positions 
are very similar to each other, their strengths are different. These
structural differences within the densities can be measured
via the {\it average} shift of the peak position along the $b_x$
direction as follows:  
\begin{equation}
\label{eq:SHIFT}
\langle b^{\phi,f}_{y}\rangle=\frac{\int
  d^2\,b_\perp\,b_{y}\,\rho^{\phi,f}_1(b_\perp,s_\perp)}{\int
  d^2\,b_\perp\,\rho^{\phi,f}_1(b_\perp,s_\perp)}  
=\frac{1}{2m_{\phi}}\frac{B^{\phi,f}_{10}(0)}{A^{\phi,f}_{10}(0)}.
\end{equation}
Using this, we obtain $\langle b^{\phi,u}_{y}\rangle=0.168$ fm and
$\langle b^{\phi,s}_{y}\rangle=0.166$ fm for model I, and $\langle
b^{\phi,u}_{y}\rangle=0.133$ fm and $\langle
b^{\phi,s}_{y}\rangle=0.100$ fm for model II.  In the
previous work, using Eq.~(\ref{eq:SHIFT}), we obtain $\langle
b^\pi_{y}\rangle=0.161$ fm, which is compatible with the lattice
calculation $\langle b^\pi_{y}\rangle=(0.151\pm0.024)$
fm, being extrapolated to $m_\pi=140$ MeV~\cite{Brommel:2007xd}, which
is similar to that of the kaon from model I. However, it turns out
that the result for the kaon is about $30\%$ smaller than that for the
pion from model II. In other words, if one considers the
phenomenologically preferable results, i.e. model II, the difference
in the polarized densities for the pion and kaon becomes more obvious,
which implies that the polarization effect of the spin inside the pion
gets more evident than inside the kaon. 

Anticipating the results from lattice QCD in near future, we  
present the relevant numerical results at $\mu = 2$ GeV which
is a usual scale of the lattice simulation. For this purpose, we want
to take into account the renormalization-group (RG)
evolution~\cite{Barone:2001sp,Broniowski:2010nt} as follows: 
\begin{equation}
\label{eq:evolve}
B_{n0} (Q^2,\mu)=B_{n0}(Q^2,\mu_0) 
\left[\frac{\alpha(\mu)}{\alpha(\mu_0)}\right]^{{\gamma_{n}}/{(2\beta_0)}},        
\end{equation}
where we have used the anomalous dimensions $\gamma_1=8/3$ and
$\gamma_{2} =8$, and $\beta_0= 11N_c/3 - 2N_f/3$ ($N_c=3$ and $N_f=3$
in the present case).  Thus, the powers in the LO evolution equation
are given as $4/27$ and $4/9$, respectively, for $n=1$ and $n=2$,
which indicate that the dependence of the tensor 
charge on the normalization point turns out to be rather weak. Note
that the anomalous dimension is simply the same as that for the
nucleon tensor charge~\cite{Kim:1995bq}. We also take
$\Lambda_{\mathrm{QCD}}=0.248\,\mathrm{GeV}$  which was    
also used in evolving the nucleon tensor charges and anomalous
magnetic moments~\cite{Ledwig:2010tu,Ledwig:2010zq}. Since the
normalization point of the present model is around
$0.6\,\mathrm{GeV}$, whereas the lattice calculation was carried out at
$\mu=2\,\mathrm{GeV}$, the scale factors turn out to be 
\begin{equation}
\label{eq:RG}
B^{\phi,f}_{10} (Q^2,\mu=2\,\mathrm{GeV})
=0.89\,B^{\phi,f}_{10} (Q^2,\mu_0=0.6\,\mathrm{GeV}) ,
\,\,\,\,
B^{\phi,f}_{20} (Q^2,\mu=2\,\mathrm{GeV})
=0.70\,B^{\phi,f}_{20}(Q^2,\mu_0=0.6\,\mathrm{GeV}). 
\end{equation}
The corresponding results for the tensor charges at $\mu=2$ GeV are  
listed in Table~\ref{TABLE2}. 

\begin{table}[htb]
\begin{tabular}{c|c|c|c|c|c|c|c} \hline\hline
\multicolumn{3}{c|}{Model I}&\multicolumn{3}{c|}{Model II}
&\multicolumn{2}{c}{Pion~\cite{Nam:2010pt}}\\
\hline
Quark
&\,\,\,\,$B^{K,f}_{10}(0)$\,\,\,\,
&\,\,\,\,$B^{K,f}_{20}(0)$\,\,\,\,
&Quark
&\,\,\,\,$B^{K,f}_{10}(0)\,\,\,\,$
&\,\,\,\,$B^{K,f}_{20}(0)$\,\,\,\,
&\,\,\,\,$B^{\pi,u}_{10}(0)$\,\,\,\,
&\,\,\,\,$B^{\pi,u}_{20}(0)$\,\,\,\,\\
\hline
$u$&$0.783$&$0.139$&$u$&$0.611$&$0.090$&
\multirow{2}{*}{$0.217$}&\multirow{2}{*}{$0.034$}\\
$s$&$-0.676$&$-0.100$&$s$&$-0.421$&$-0.076$
&&\\
\hline \hline
\end{tabular}
\caption{Renormalization-group (RG)  evolution of the results for the
  $B^{K,q}_{10}(0)$ and $B^{K,q}_{20}(0)$ at the scale $\mu=2$ GeV
  with $m_K=495$ MeV for model I and model II. We also write the
  results for the pion.}  
\label{TABLE2}
\end{table}
\vspace{0.5cm}

\textbf{5.} In the present work, we aimed at investigating the spin
structure of the kaon, based on the nonlocal chiral quark model from
the instanton vacuum. We first evaluated the generalized form factors for the
kaon, i.e. vector and tensor form factors for the moments $n=1,2$. 
We calculated the flavor vector and tensor form factors of the kaon
with explicit flavor SU(3) symmetry breaking considered.  
In order to improve the electromagnetic form factors of the kaon,
which was studied previously~\cite{Nam:2007gf}, we treated the
constituent-quark mass at zero virtuality of the quark as a free
parameter to fit the experimental data. The vector properties of the
pion were almost not changed, whereas those of the kaon were shown to
be much improved by using the higher value of the constituent-quark
mass. Both the results for the pion and kaon  were in good agreement
with the data.  

Having evaluated the generalized vector and tensor form factors of the
kaon, we proceeded to compute the probability densities of the
polarized quark inside the kaon. In doing so, we parametrized the form
factors, employing the simple monopole and $p$-pole type
parameterizations for the vector and tensor ones. Using 
the parametrized results, we computed the probability densities of the
unpolarized and polarized quarks inside the kaon as functions of the
impact parameters, which reveal the spin structures of the kaon.
Anticipating the data from the lattice simulation in near future, 
we also presented the results of the tensor charges, evolving them 
from the present scale $\mu_0\approx600$ MeV to $\mu=2$ GeV which is a 
scale often used in the lattice simulations. 

We summarize the important observations in the present work: 
\begin{itemize}
\item The electromagnetic form factor of the pion is reproduced
  quantitatively well for both models, whereas that of the kaon 
  is well described within model II, i.e. the phenomenological
  one, compared to the original model (model I) in which there is no
  free parameter. This difference can be understood by the absence of
  the $1/N_c$ meson-loop corrections in the present work, which will
  improve the results of the original model.   
\item Due to the explicit flavor SU(3) symmetry breaking, the up and
  strange form factors turn out to be asymmetric with respect to the
  interchange of up and strange quarks, i.e. $V$-spin transformation,
  being different from the pion case. In general we find that the
  strange form factors are relatively flat in comparison to the up
  form factors.  
\item In parametrizing the form factors, we find that model II
  provides larger pole masses  $(0.894-1.520)$ GeV than those from
  model I  $(0.647-0.806)$ GeV. Note that the results of model I are
  closer to those of the pion.  
\item Considering the tensor form factors, we find that the
  probability density for the polarized quarks get distorted. The degree of
  this distortion can be measured by the deviation of the average
  value of $\langle b^{K,f}_y\rangle$ from zero. In model II, the
  deviation depending on quark species is seen more obviously,
  i.e. $\langle  b^{K,(u,s)}_y\rangle=(0.133,0.100)$ fm, while almost
  no difference is observed for model I. 
\item The RG evolution of the the present results brings about the
  tensor charges as follows: $B^{K,u}_{10}=(0.611-0.783)$,
  $|B^{K,s}_{10}|=(0.421-0.676)$, $B^{K,u}_{20}=(0.090-0.139)$,
  and $|B^{K,s}_{20}|=(0.076-0.100)$ at the renormalization scale
  $\mu=2$ GeV, theoretical uncertainties of the present model being
  taken into account. 
\end{itemize}

As we have shown in the present work, the generalized form factors
play a role of revealing the internal spin structures of mesons. While it
is very difficult to get access directly to the spin structures of
mesons, the lattice data will shed light on understanding them. 
Moreover, the generalized form factors with different operators and
with higher moments will further show us how the quarks inside the
kaon behave more in detail.  Thus, it will be of great interest to
investigate them in the future.  

\section*{Acknowledgments}
The authors are grateful to B.~G.~Yu and  Gh.~-S.~Yang for fruitful
discussions. S.i.N. is thankful to the hospitality of the Hadron
Theory Group at Inha University during his visit, where this work was
performed. The work of H.Ch.K. was supported by Basic Science Research
Program through the National Research Foundation of Korea (NRF) funded
by the Ministry of Education, Science and Technology (grant number:
2010-0016265). The work of S.i.N. was supported by the grant
NRF-2010-0013279 from National Research Foundation (NRF) of Korea.  

\end{document}